\documentstyle[12pt]{article}
\begin{document}
\topmargin -1.4cm
\oddsidemargin -0.8cm
\evensidemargin -0.8cm

\title{A center vortex representation of\\ the classical SU(2) vacuum}

\vspace{1.5cm}

\author{P. Olesen\\
{\it   The Niels Bohr Institute, University of Copenhagen,}\\
{\it Blegdamsvej 17, Copenhagen \O, Denmark} }

\maketitle
%\vfill
\begin{abstract}
The classical massless SU(2) field theory has an infinite number
of gauge equivalent representations of the vacuum. We show that among these
there exists a non-perturbative center vortex representation with 
some similarity to the
quantum vacuum of the same theory. This classical SU(2) vacuum 
consists of a lattice of center vortex pairs combining to triviality. 
However,this triviality can be broken by perturbations, for example by adding 
a mass term, or considering the electroweak theory where the Higgs field
does the breaking, or by quantum fluctuations like in QCD. 

\end{abstract}

\thispagestyle{empty}

\vskip0.5cm

In non-Abelian field theories the physical vacuum can be considerably 
complicated. For example, in the quantum QCD vacuum a condensate of
center vortices is expected.  Recent discussions of the SU(2) case can be 
found in ref. \cite{tyskere} and \cite{jeff}. In the following we shall 
show that something similar is the 
case even for the classical vacuum in a certain non-perturbative 
representation. 
The difference between the two cases is that classically there is no 
scale parameter, so all scales are arbitrary, in contrast to the quantum case.
Also, in the classical case the vacuum consists of pairs of non-trivial center
vortices so the net effect in SU(2) is a flux $(-1)^2$, which gives the 
classical vacuum the usual trivial appearence. However, the fact that this
appearence can be understood from a center vortex point of view 
makes it perhaps more natural that this kind of vortex lattice also occurs in
the quantum state. In the quantum case the necessary scale for the transverse 
size of each vortex is provided by quantum mechanics.

In this note we investigate a periodic vacuum (zero energy) solution of the 
classical massless
SU(2) Yang Mills theory.  The SU(2) field strength
is thus assumed to vanish
($A_\mu=A_\mu^a~t_a,~t_a=\sigma_a/2$ ),
\begin{equation}
F_{\mu\nu}=\partial_\mu A_\nu-\partial_\nu A_\mu+ig[A_\mu,A_\nu]=0.
\end{equation}

The trivial vacuum field 
$A_\mu=0$ of course satisfies this. Any other vacuum configuration must be of 
the form
\begin{equation}
A_\mu=A_\mu^at_a=\frac{i}{g}~\partial_\mu\Omega~\Omega^\dagger\label{omega}.
\end{equation}
Alternatively the unitary matrix $\Omega$ can be expressed in terms of the field
\begin{equation}
\Omega={\rm P}~\exp \left(-ig\int _R^x A_\mu~dx^\mu\right),
\end{equation}
where $R$ is some arbitrary reference point.
Our ansatz for the vacuum solution is based on the fields
\begin{equation}
A_\mu^3 ~{\rm with} ~\mu=1,2~~{\rm and}~~W_\mu=\frac{1}{\sqrt{2}}\left(A_\mu^1+
iA_\mu^2 \right), ~W_2=iW_1\equiv iW,~W_3=W_0=0.
\end{equation}
We assume periodicity in the $x_1-x_2$ plane.
The field $W$ stabilizes the dynamics, since if it was not present, an
instability would
be generated, at least  for suffiently homogeneous fields \cite{nkn}.
Then the solution of $F_{\mu\nu}=0$ for the $A^3$-field satisfies the first order
equations
\begin{equation}
(D_1+iD_2)W=0,~~f_{12}=\partial_1A_2^3-\partial_2A_1^3=2g^2~|W|^2.~~D_i=
\partial_i-igA_i^3\label{firstorder}
\end{equation}
Similar equations were used long time ago in the massive SU(2) case \cite{ja}. 
It is possible to show directly from (\ref{firstorder}) that the second order 
equations of motion are satisfied by use of
\begin{equation}
(D_1-iD_2)(D_1+iD_2)W=0,~{\rm and}~[D_1,D_2]=-igf_{12}
\end{equation}
which follows from the first equation (\ref{firstorder}). Hence
\begin{equation} 
(D_1^2+D_2^2+2g f_{12})W-2g^2|W|^2W=0,
\end{equation}
which is precisely one of the equations of motion for our ansatz. The second 
equation is derived by simply differntiating the second equation in
(\ref{firstorder}),
\begin{equation}
\partial_i f_{ij}=2g^2\epsilon_{ij}\partial_i|W|^2,
\end{equation}
showing that the magnetic field is generated by a current from the complex
vector fields.
 
The equations (\ref{firstorder}) can be reexpressed as
\begin{equation}
A_i^3=\frac{\epsilon_{ij}}{g}\partial_j\log|W|+\frac{1}{g}\partial_i\chi,
\label{A3}
\end{equation}
where $\chi$ is the phase of $W$,
and $W$ satisfies the Liouville equation,
\begin{equation}
-(\partial_1^2+\partial_2^2)\ln |W|=2g^2~|W|^2-\epsilon_{ij}\partial_i\partial_j
\chi.
\label{liouville}
\end{equation}
These equations are non-perturbative. The magnetic field and the $W-$field
are in a bootstrap situation: The field $f_{12}$ is generated by a current 
arizing from the charged $W-$field, and the latter appears in order to 
stabilize the magnetic field which would otherwise be unstable as 
dicussed in \cite{nkn}.

To proceed we take for simplicity the periodic lattice to consist of
quadratic cells  $\omega\times i\omega$ and a solution which has a
non-trivial topology
\begin{equation}
W(z,\bar{z})=\frac{\sqrt{2}}{g}~\frac{|e_1||\wp' (z)|}{|e_1|^2+|\wp (z)|^2}
~e^{i\chi},~~\chi=\sum_i {\rm arg}(z-z_i),~~z_i=\omega n+i\omega m
\label{solution}
\end{equation}
can be obtained.
Here $z_i$ are the first order zeros of $W$ encircled by the phase $\chi$.
Also, $\wp$ is the doubly periodic Weierstrass function with periods 
$2\omega,~2i\omega$.
The solution (\ref{solution}) has, however,  periods $\omega,~i\omega$ 
\cite{poul91}. The constant $e_1$ known in the theory of the
Weierstrass function is carefully arranged\footnote{More general versions of 
the solution (\ref{solution}) have been given in \cite{hollaendere}} in 
Eq. (\ref{solution})
such that in one cell one only has one zero of $W$. In general a construction 
in terms of Weierstrass' function (or any other elliptic function)
a la (\ref{solution}) leads to an even number of zeros. The flux
would then be trivial\footnote{This is because the flux $\int f_{12}~ d^2x$
through one cell would then be an even number times $2\pi$. Since
the field is associated with $\sigma_3/2$ in SU(2), this gives 0, 2$\pi$ etc.
for the encircling angle. With only one zero the corresponding angle
is $\pi$,3$\pi$, etc. } in SU(2).

Next we shall evaluate the Wilson loop taken along the sides of
a fundamental lattice cell.   The
corners are placed at $C_1=-\omega/2-i\omega/2,~C_2=+\omega/2-i\omega/2,~
C_3=+\omega/2+i\omega/2,$ and
$C_4=-\omega/2+i\omega/2$. We now make a transformation $\Omega$ of the field 
$A_\mu=A^a_\mu t_a$ along this loop,
\begin{equation}
{\hat A}_\mu =\Omega A_\mu \Omega^\dagger-\frac{i}{g}\Omega\partial_\mu 
\Omega^\dagger,
\end{equation}
where
\begin{equation}
\Omega=
\left\{
\begin{array}{cc}
e^{i\chi /2} & 0 \\
0 & e^{-i\chi/2} \\
\end{array}
\right\}=e^{i\chi t_3}
\label{omega4}
\end{equation} 
This transformation 
accomplishes the ``removal'' of the gradient term in the field
$A^3_i$ in Eq. (\ref{A3}). The new fields are given by
\begin{equation}
\hat{A}_i^3=\frac{\epsilon_{ij}}{g}\partial_j\log|W|,~W_1=|W|,~W_2=i|W|. 
\label{hatA3}
\end{equation}
Therefore the phase $\chi$ has been transformed away from both $A^3_i$ and the
$W$-fields. We note that for our ansatz
\begin{equation}
A_1^1t_1+A_1^2t_2=\left\{
\begin{array}{cc}
0 & \sqrt{2}W^\star \\
\sqrt{2}W & 0 \\
\end{array}
\right\}~~{\rm and}~~A_2^1t_1+A_2^2t_2=\left\{
\begin{array}{cc}
0 & -i\sqrt{2}W^\star \\
i\sqrt{2}W & 0 \\
\end{array}
\right\}.
\end{equation}
Therefore with $\Omega$ given as in Eq. (\ref{omega4}) we obtain
\begin{equation}
\Omega (A_1^1t_1+A_1^2t_2)\Omega^\dagger =\sqrt{2}|W| t_1
\end{equation}
and similarly for the 2 components, so to sum up
\begin{equation}
\hat{A}_1=\hat{A}_1^3~t_3+\sqrt{2}~|W|~t_1, ~\hat{A}_2=\hat{A}_2^3~t_3+\sqrt{2}~
|W|~t_2.
\label{hatA}
\end{equation}
It should be remarked that this transformation would be bad near the zeros of
$W$, because the first term on the right hand side of Eq. (\ref{A3}) is
 singular at a zero, but this is exactly canceled by the gradient term
in this equation, making the $A^3$-fields finite at the zeros. However, this
problem does not occur along the contour ${\cal C}=C_1-C_2-C_3-C_4-C_1$,
where the new field ${\hat A}^3$ is perfectly finite.

Along the contour $\cal C$  the fields are simplified
 in an essential manner. Thus, along $\cal C$ the field $|W|$ has  maximum and
no slope in the direction transverse to this contour.
 Therefore $\partial_2 \log|W|=0$ along the lines $C_1-C_2$ and
$C_3-C_4$, and  $\partial_1 \log |W|=0$ along the lines $C_2-C_3$ 
and $C_4-C_1$.
Therefore it follows from
Eqs. (\ref{hatA3}) that along the $C_1-C_2$ and $C_3-C_4$ lines
\begin{equation}
\hat{A}_1^3=0~{\rm on }~C_1-C_2~{\rm and}~  C_3-C_4.
\end{equation}
Similarly
\begin{equation}
\hat{A}_2^3=0~{\rm on }~C_2-C_3~{\rm and}~  C_4-C_1.
\end{equation}
It is therefore a consequence that  the $\hat{A}$ field only 
has contributions from
$|W|$, as is seen from Eq. (\ref{hatA}), 
\begin{equation}
\hat{A}_1=\sqrt{2}|W|~t_1~{\rm on}~C_1-C_2~{\rm and}~  C_3-C_4,
\label{e7}
\end{equation}
and
\begin{equation}
\hat{A}_2=\sqrt{2}|W|~t_2~{\rm on}~C_2-C_3~{\rm and}~  C_4-C_1,
\label{simpel}
\end{equation}
This is the important simplification which allows us to compute the Wilson
loop around the boundary of the fundamental cell ${\cal C}=C_1-C_2-C_3-C_4-
C_1$.

We have
\begin{equation}
W({\cal C})={\rm tr~P}~\exp \left(ig\int_{\cal C }A_\mu dx_\mu\right)=
{\rm tr~P}~\left[\Omega^\dagger _{\rm initial}\exp \left(ig\int_{\cal C} 
\hat{A}_\mu dx_\mu\right)~\Omega_{\rm final}\right].
\label{e9}
\end{equation}
Now $\Omega_{\rm final}$ differs from $\Omega_{\rm initial}$ by the center 
element (-1). Therefore
\begin{equation}
W({\cal C})=(-1)_{\rm A^3}{\rm tr~P}~\exp \left(ig\int_{\cal C }
\hat{A}_\mu dx_\mu\right).
\label{e10}
\end{equation}
We have written the (-1) in this special way  in order to remind us that
this center contribution comes from the original field $A_3$.

Next  point is that the integral along the different paths is the
same, due to the fact that the function $|W|$ 
 is symmetric in  $x_1$ and $x_2$, so we have
\begin{eqnarray}
&&g\int_{c_1-c_2}\hat{A_1}dx_1=g\int_{c_2-c_3}\hat{A_2}dx_2\equiv 2I  
\nonumber\\
&&g\int_{c_3-c_4}\hat{A_1}dx_1=g\int_{c_4-c_5}\hat{A_2}dx_2\equiv -2I,~I=
\frac{g}{\sqrt{2}}\int_{C_1-C_2}|W(x_1-i\omega/2)|~dx_1.
\label{e11}
\end{eqnarray}
This is easily seen because the function $|W|$ only depends on $x_1+ix_2$, and 
therefore the integrations along $C_1-C_2$ and $C_2-C_3$ etc. produce the same
results. The minus signs simply arise from the inversion of the paths
of integration, taking into account the periodicity of $|W|$.

Collecting these results we obtain
\begin{equation}
W({\cal C})=(-1)_{\rm A^3}{\rm tr}~\left[e^{i\sigma_1 I}e^{i\sigma_2 I}
e^{-i\sigma_1 I}e^{-i\sigma_2 I}\right]
\label{e12}
\end{equation}
By use of the  relation $e^{i\sigma_1~ I}=\cos I+i\sigma_1~\sin I$, etc.,
we easily obtain
\begin{equation}
W({\cal C})=(-1)_{\rm A^3}(1-2\sin^4(I))_{W},
\label{13}
\end{equation}
where we used the commutator
\begin{equation}
e^{i\sigma_2~ I}e^{-i\sigma_1~ I}-e^{-i\sigma_1~ I} e^{i\sigma_2~ I}=
[\sigma_2,\sigma_1]\sin^2[I]=-2i\sin^2[I]~\sigma_3.
\end{equation}
The index $W$ on $1-2\sin^4(I)$ is there to remind us that this
contribution comes from the $W$-field.

The integral $I$ is explicitly given by
\begin{equation}
I=\int_{-\omega/2}^{\omega/2}~dx_1~\frac{e_1 |\wp' (x_1-i\omega/2)|}
{e_1^2+|\wp (x_1-i\omega/2)|^2}.
\end{equation}
Scaling the complex variable by $\omega$ this integral can be written
\begin{equation}
I=\int_{-1/2}^{1/2}~du_1~\sqrt{c}\frac{|\wp' (u_1-i/2)|}
{c+|\wp (u_1-i/2)|^2},~~c=(e_1\omega^2)^2=\frac{g_2}{4}\omega^4=15\sum_{mn} 
\frac{1}{(2n+2im)^4},
\end{equation}
where $m,n$ are different from (0,0). The Weierstrass function above is
periodic in the cell $2+i2$.
The constant $c$ is therefore independent of $\omega$, so $I$ is independent of
$\omega$, and we only need to 
evaluate $I$ once. I have not been able to compute
this integral analytically, but a high precision numerical integration  
gives\footnote{By use of WolframAlpha on a smartphone I obtained
$I-\pi/2=2.157\times 10^{-6}$, where $g_2=11.8171$}
\begin{equation}
I=\frac{\pi}{2}.
\end{equation}
Inserting this in (\ref{13}) we obtain the result
\begin{equation}
W({\cal C})=(-1)_{\rm A^3}(-1)_{W}
\end{equation}
The result is therefore that the Wilson loop gets a center vortex 
contribution from the $A^3$-field and another such contribution from the 
$W$-field. Since $(-1)^2=+1$
the contribution from this pair of center vortices is the trivial
unit element, corresponding to the natural expectation that the {\it classical}
vacuum 
does not carry a magnetic flux, in accordance with the possibility of
gauge transforming all fields $A_\mu$ to zero. A similar conclusion
can be obtained in a simpler manner by using that when the field
strength $F_{\mu\nu}$ vanishes, the Wilson loop can be shown to be independent 
of the loop \cite{yuri} when there are no singularities. Therefore the 
contour can be contracted to a
small circle around the zero, and again we get two contributions which cancel.

In the case of the electroweak SU(2)$\times$U(1) theory there is a
similar magnetic vortex lattice representation of the vacuum in the
symmetric phase. This can be verified using the same methods as above.
We also mention that our approach may have applications for
Chern Simons dynamics.

The conclusion is thus that the classical SU(2) vacuum is a sea of pairs of 
non-trivial center vortices which combine to triviality. Perturbations actually 
converts the magnetic flux so as to become physical. For example, adding a 
mass term the
center vortex state appears in a non-trivial \cite{ja} manner. Here the 
Liouville
equation (\ref{liouville}) is replaced by
\begin{equation}
-(\partial_1^2+\partial_2^2)\ln |W|=m^2+2g^2~|W|^2-\epsilon_{ij}\partial_i
\partial_j\chi,
\label{nw}
\end{equation} 
where $m$ is a mass added to the Lagrangian by a term $-m^2W_\mu W_\nu^\dagger$.
The resulting vortices now have a scale set by the mass $m$.
A similar result  is true in the 
electroweak SU(2)$\times$U(1) theory, where the Higgs field breaks the 
triviality \cite{poul91}. This is directly related to a phase transition
from the massless to the massive case.

For the case of the quantum fluctuations the situation was discussed by 
Ambj\o rn and the author \cite{jan}, where we considered the Savvidy
magnetic field $H_S$ \cite{savvidy} as a background field. The Liouville 
equation is then replaced by
\begin{equation}
-(\partial_1^2+\partial_2^2)\ln |W|=gH_S-\epsilon_{ij}\partial_i
\partial_j\chi,
\label{cc}
\end{equation}
which is the similar to Eq. (\ref{nw}) with the mass replaced by the Savvidy 
field $H_S$ and also assuming that $W$ is much smaller than $H_S$. This
assumption makes it possible to ignore the $2g^2 |W|^2$ term which in
general would occur on the right hand side like in Eq. (\ref{nw}). The 
solution of Eq. (\ref{cc}) 
is different
from Eq. (\ref{solution}). An example of a solution is \cite{jan}
\begin{equation}
W(z)={\rm const.}~e^{-gH_S x^2/2}~\theta (z),
\end{equation}
where $\theta$ is a theta function. For a detailed discussion we refer to
ref. \cite{jan}. The integral which replaces $I$ is now much smaller than
$\pi/2$ because $W$ is small, so in the resulting Wilson loop the
$(-1)_{\rm A^3}$ contribution cannot be overwhelmed by the $1-2\sin^4 I$ factor 
in Eq. (\ref{13}). We therefore get a non thrivial SU(2) center vortex.
\vskip0.5cm

I thank Michael Engelhardt for several enlightening and interesting discussions.

\end{document}